\documentclass[a4paper,fleqn]{cas-dc}

\usepackage[authoryear,longnamesfirst]{natbib}
\usepackage{placeins}
\usepackage{graphicx}
\usepackage{siunitx}
\usepackage{hyperref}
\usepackage{subcaption} % For subfigures
\usepackage{caption}
\def\tsc#1{\csdef{#1}{\textsc{\lowercase{#1}}\xspace}}
\tsc{WGM}
\tsc{QE}
\tsc{EP}
\tsc{PMS}
\tsc{BEC}
\tsc{DE}
\newcommand{\hfigref}[1]{\hyperref[#1]{Fig.~\ref*{#1}}}
\begin{document}
\let\WriteBookmarks\relax
\def\floatpagepagefraction{1}
\def\textpagefraction{.001}
\shorttitle{Controlling the Coherence at CLS using X-ray Interferometry}
\shortauthors{Y.~Y.~Sigari et~al.}
%\begin{frontmatter}

\title [mode = title]{Controlling \& Measuring the Degree of Coherence at CLS using X-ray Interferometry}                      
\tnotemark[1]

\tnotetext[1]{This document is the results of the research project funded by the NSERC.}
%\tnotetext[2]{}
\author[1, 2]{Y.~Y.~Sigari}[type=editor,
                        %auid=000,bioid=1,
                        prefix=,
                        role=Researcher,
                        orcid=0000-0002-3736-1905
                        ]
\cormark[1]

\ead{yysigari@triumf.ca}
\affiliation[1]{organization={TRIUMF},
                addressline={4004 Wesbrook Mall}, 
                city={Vancouver},
%               citysep={}, % Uncomment if no comma needed between city and postcode
                postcode={V6T 2A3}, 
                state={BC},
                country={Canada}}
\affiliation[2]{organization={Department of Physics \& Eng. Physics, University of Saskatchewan},
                addressline={105 Administration Pl}, 
                city={Saskatoon},
%               citysep={}, % Uncomment if no comma needed between city and postcode
                postcode={S7N 5E2}, 
                state={SK},
                country={Canada}}

%\credit{Experiment}
\author[2]{N.~A.~Simonson}[%
   role=Co-researcher,
   %suffix=,
   ]
\author[3]{N.~Appathurai}[%
role=Beamline Responsible]
\author[2]{R. Castle}[%
   role=Co-researcher,
   %suffix=,
   ]
\author[3]{B.~D.~Moreno.}[
            role=Beamline Responsible]
\author[3]{S.~Saadat}[
            role=Accelerator Division] 
\author[3]{J.~Wang}[
            role=Beamline Responsible] 
%\fnmark[2]
%\ead{wjh@example.org}
%\ead[URL]{https://www.university.org}
%\credit{Data curation, Writing - Original draft preparation}
\affiliation[3]{organization={Canadian Light Source},
                addressline={44 Innovation Boulevard}, 
                city={Saskatoon},
                %postcodesep={}, 
                postcode={S7N 2V3}, 
                state={SK},
                country={Canada}}
\author[3]{J.~M.~Vogt}[%
   role=Co-supervisor,
   %prefix=Dr,
   ]
\author[2, 3]{M.~J.~Boland}[%
   role=Supervisor,
   %prefix=Dr.,
   ]
%\cormark[2]
%\fnmark[1,3]
%\ead{mark.boland@usask.ca}
%\ead[URL]{www.campus.in}
\cortext[cor1]{Principal corresponding author}
%\cortext[cor2]{Principal corresponding author}
\fntext[fn1]{The data is uploaded under doi:10.5281/zenodo.16416212, for public access after the publication.}
%\nonumnote{This note has no numbers. In this work we demonstrate $a_b$ }

\begin{abstract}
This paper investigates a case study on measuring and controlling the first-order degree of spatial coherence under different coupling adjustments in the storage ring.\
The experimental findings are consistent with the predicted inverse relationship between the visibility and the coupling factor.\
The degree of coherence was measured using X-ray double slit interferometry with synchrotron radiation at an energy of~\SI{7}{keV} on the Brockhouse X-Ray Diffraction and Scattering in-vacuum undulator beamline.\
The vertical degree of coherence increases as the coupling factor in the storage ring is reduced.\
The Linear Optics for Closed Orbit (LOCO) algorithm is used to model the linear terms of the storage ring optics in Accelerator Toolbox.\
The LOCO-tuned model provides insights into the variations in the vertical beam size at two different source points in the storage ring as a function of the coupling factor.\
The coupling factor is parameterized by the closest-tune approach with a bunch-by-bunch feedback system to confirm the trend in the changes of the vertical beam size and the visibility.\

\end{abstract}
%\begin{graphicalabstract}
%\includegraphics{figs/cas-grabs.pdf}
%\end{graphicalabstract}
%\begin{highlights}
%\item Research highlights item 1
%\item Research highlights item 2
%\item Research highlights item 3
%\end{highlights}

\maketitle
\section{Introduction}
Synchrotron light sources are state-of-the-art research facilities that provide synchrotron radiation with exceptional spectral brightness.\
Advancements in light sources coherence properties pave the way for future research areas~\cite{Shin2021}.\
The enhanced spectral brightness enabled the development of leading-edge research techniques such as hierarchical phase-contrast tomography~\cite{cdi_gale_infotracmisc_A769864464}.\
As light source facilities evolve, the monitoring instrumentation and techniques must keep pace with rapid design changes and operational challenges~\cite{4thgen_chal_2014}.\
Fourth generation light sources benefit from the high degree of coherence provided by the multi-bend-achromat lattices and the small transverse emittance of the electron beam.\
Although the current third generation light source cannot match this quality, one still can extend their limits by using extra focusing effects provided by the skew quadrupoles in the storage ring lattice, and users have shown interest in overcoming these limitations~\cite{Takayama:el3109, Chushkin:ro5044, castle2025}.\
X-ray double-slit interferometry, developed as an emittance-monitoring tool for fourth-generation light sources and Future Circle collider, shows great promise for measuring the transverse beam size of a diffraction-limited emittance storage ring~\cite{mitsuhashi_lhc_vdbl, Mitsuhashi2016}.\
By measuring the spatial degree of coherence of a randomly distributed source and applying the van~Cittert-Zernike theorem, X-ray interferometry allows reconstruction of the intensity profile of synchrotron radiation source point~\cite{Born_Wolf}.\
Beyond source size monitoring, this technique offers the potential to enable users  to control the degree of coherence of synchrotron radiation by adjusting the storage ring lattice parameters.\

This paper presents the interferometry data collected at the Brockhouse X-ray Diffraction and Scattering Sector (BXDS-IVU) beamline~\cite{bxds_diaz_2014} at the Canadian Light Source (CLS)~\cite{canlight_CutlerJeffrey2017Tbli, cls_design_dallin_2003}.\
The study shows how to improve the first-order spatial coherence by decreasing the coupling of the electron beam between vertical and horizontal axis of the electron beam dynamics; this is achieved through changing the magnetic strength of a skew quadrupole in the CLS Double Bend Achromat (DBA) cell~\cite{Couple_WURTZ_2018}.\
A model of the CLS storage ring in Accelerator Toolbox was prepared with the Linear Optics for Closed Orbit (LOCO) algorithm to simulate the changes of the vertical beam size as a function of coupling~\cite{loco_safranek2009, canmag_dallin_2003, YousefiSigari:2021hwg}.

\section{Theory}\label{sec:theory}
\subsection{Interferometry}
In a storage ring, the electron beam consists of randomly distributed synchrotron radiation source points; the emitted radiation from each point has a phase that is statistically independent from the others.\
The intensity profile is an indicator of the correlation between the random phases, also called the degree of coherence, which is experimentally measured as the maximum visibility of the interferograms~\cite{vanCittert1934, Zernike1938, Born_Wolf}.\
In X-ray double-slit experiment, the synchrotron radiation diffracts from the two slits, made parallel on a foil, and forms an interference pattern on a scintillation detector.\
The intensity profile on the scintillator is governed by the following equation
\begin{equation}\label{eq:1}
\begin{aligned}
I_{\text{observed}}(y)
&= I_0 \,\text{sinc}^2\!\left(\frac{2\pi a}{\bar{\lambda} R}\, y\right) \\
&\quad \times \left[ 1 + \lvert \gamma(D) \rvert
\cos\!\left(\frac{\pi D}{\bar{\lambda} L}\, y + \phi \right) \right],
\end{aligned}
\end{equation}
where $I_0$ is the maximum intensity, $y$ is the vertical distance from the center of the interference pattern in the detector plane, $a$ is width of each slit, $R$ is the optical distance from the double slit foil to the scintillator, $\bar{\lambda}$ is the average wavelength of the radiation, $D$ is the space between the centers of the two parallel slits, $\phi$ is the phase offset, and $|\gamma(D)|$ is the modulus of the first-order degree of spatial coherence~\cite{Born_Wolf, young-1804-interference, mitsuhashi_lhc_vdbl}.\
For a double slit experiment, the spatial degree of coherence $\gamma(D)$ is measured as a function of slit spacing $D$.\
In experiments~\cite{Born_Wolf}, the $|\gamma(D)|$ is equivalent to the fringe visibility between the maximum and minimum intensities $I_\text{max}$ and $I_\text{min}$ given by
\begin{equation}\label{eq:3}
    \text{Visibility} = \frac{I_{\text{max}}-I_{\text{min}}}{I_{\text{max}}+I_{\text{min}}}.
\end{equation}
\subsection{Coupling} The vertical beam emittance primarily arises from the coupling of the vertical beam dynamics to the horizontal emittance as the inherent vertical emittance, $\epsilon_y$, is very small compared to the horizontal emittance, $\epsilon_x$.\
This relationship between the vertical emittance and coupling, $\kappa$, is given by
\begin{equation}\label{eq:6}
    \epsilon_y = \kappa \epsilon_x,
\end{equation}
where the measurement and analysis are explained in pages 56 to 62 of Ref.~\cite{minty2003}.
\subsection{Calibrated Linear Model}
\begin{figure}[!tb]
    \centering
    \includegraphics[width=\linewidth]{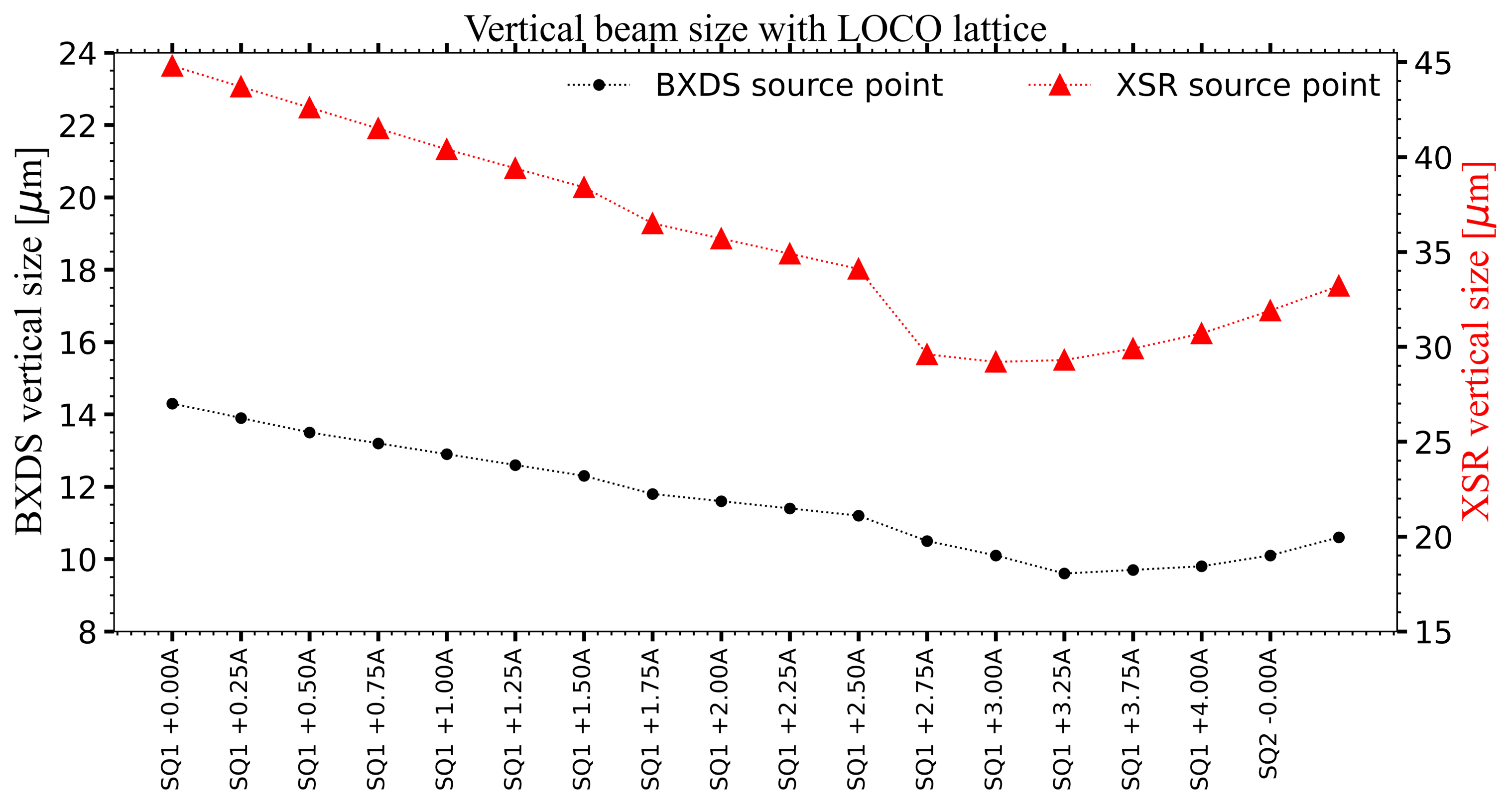}
    \caption{The calibrated model of the storage ring shows consistency between BXDS-IVU and XSR vertical source size variations in trend.}
    \label{fig:loco}
\end{figure}

A model of the storage ring was created with the Linear Optics for Closed Orbit (LOCO) algorithm~\cite{loco_safranek2009, 2021MATLABR2021a}.\
Using this model, we simulated the variation of the vertical beam size as a function of the coupling factor.\
The simulation predicts a consistent correlation between vertical beam size at the two different locations in the storage ring and the coupling.\
Both vertical beam sizes gradually decrease then increase, providing the basis for the experiment, described in section~\ref{sec:expsetup}.\
The vertical beam size, defined by the beam-envelope matrix in Ref.~\cite{bemaenvelope_ohmi_1994}, is evaluated at two beamline source points in the storage ring, XSR (red)~\cite{Bergstrom2008} and BXDS-IVU (black), as shown in Fig.~\ref{fig:loco} in Accelerator Toolbox~\cite{acctool_terebilo_2001}.
%\FloatBarrier

\begin{figure*}[!Hb]
    \centering
    \includegraphics[width=\linewidth]{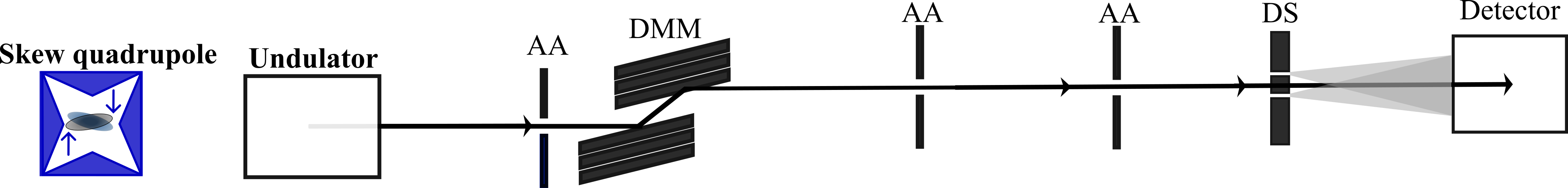}
    \caption{Schematic of the BXDS-IVU experimental layout. The skew quadrupoles in the storage ring lattice modify the transverse orientation of the electron beam in the straight section, thereby altering the degree of spatial coherence of the synchrotron radiation in the BXDS-IVU beamline. These changes are monitored using an interferometry setup.}
    \label{fig:set_up}
\end{figure*}

\section{Experimental setup}\label{sec:expsetup}
A schematic drawing of the set-up is presented in Fig.~\ref{fig:set_up}.\
The experiment involved varying the coupling factor and observing its impact on the degree of spatial coherence.\
Key components of the setup included skew quadrupoles in the storage ring, and on the BXDS-IVU beamline there were precision-engineered double slits, adjustable apertures, and a microscopic X-ray detector.\
The changes in the vertical beam size was monitored using an X-ray pinhole camera at the XSR beamline~\cite{Bergstrom2008}.\
\subsection{Physical layout}
The incoherent synchrotron radiation produced at the straight section between cell $3$ and cell $4$ passes several optical elements, including a double-multilayer monochromator (DMM), adjustable apertures (AA), the double slits (DS) mounted on a transverse motion controller, before reaching the detector~\cite{bxds_diaz_2014}.\
%\FloatBarrier

\begin{figure}[tb]
    \centering
    \includegraphics[width=\linewidth]{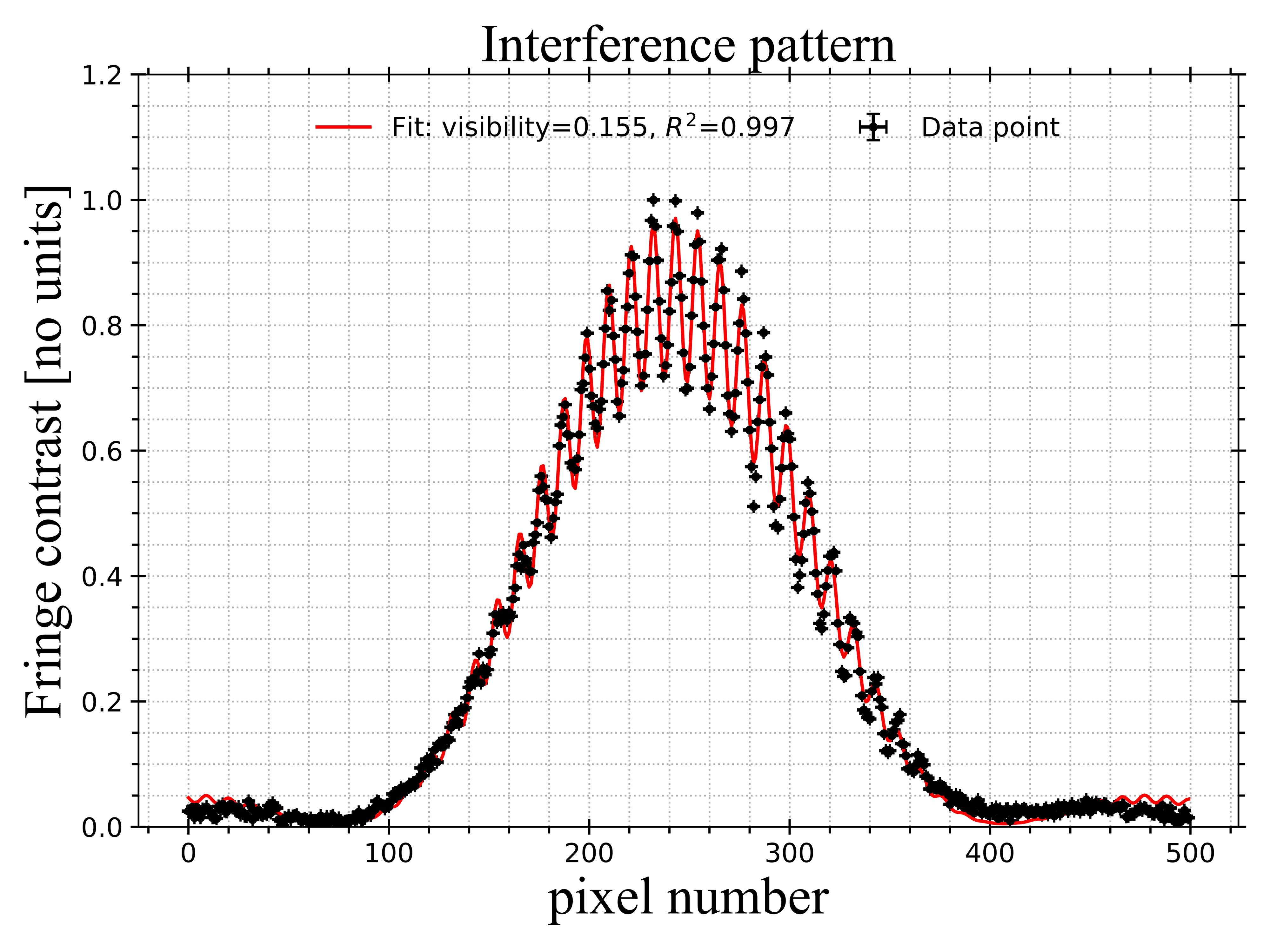}
    \caption{Interference pattern for slit spacing of \SI{50}{\micro \meter}.}
    \label{fig:d50}
\end{figure}

The outcome of each image shot is the visibility which results from processing the raw interferogram captured by the detector.\
Figure~\ref{fig:d50} represents a sample of the processed intensity profile of the interference pattern for slit spacing of~\SI{50}{\micro \meter}.\
%\FloatBarrier
%The values from Table~\ref{tbl_sring} correspond to a Gaussian FWHM of photon beam $\Sigma_{x,y}$ equal to \SI{2.175}{\milli \meter} horizontally and  \SI{0.7933}{\milli \meter} vertically at the first harmonic energy \SI{1.7}{keV} at the distance \SI{50}{\meter} from the source with coupling constant of \SI{0.35}{\%}, calculated by SPECTRA~\cite{Tanaka-specrta-2001}.
\subsection{Instruments}
\subsubsection{Skew quadrupole}
The beam coupling factor is varied using skew quadrupoles.
Each Double Bend Achromat (DBA) cell in the storage ring includes a set of sextupoles located in the middle of the cell~\cite{Chasman_1975_PreliminaryDO}.\ These sextupoles serve multiple functions, one of which is functioning as a skew quadrupole magnet.\

\subsubsection{BXDS-IVU Detector}
%The distance between the source point, the double slits and the detector puts a limit on the spatial resolution of the detector to accurately resolve the peaks of the interference pattern.\
%In this experiment, 
A 16-bit sCMOS (scientific Complementary Metal– Oxide–Semiconductor) digital camera was chosen as the detector, with a pitch size of~\SI{6.5}{\micro\meter}. 
%and a quantum efficiency of \SI{60}{\%} at \SI{600}{\nano \meter}.\
The detector is paired with monochromatic microscope optics, featuring a $LuAG$:$Ce$ scintillator of~\SI{50}{\micro\meter} thickness, which provides a fourfold magnification.\
The full detector system offers a spatial resolution of~\SI{2.1}{\micro\meter}~\cite{detector_man_2018}.\

\subsubsection{Double slits foils \& transverse motion control}
Double slits (DS) were mounted~\SI{5.5}{\meter} upstream of the detector and~\SI{50}{\meter} from the source point.\
The DS foils used in the experiment were fabricated from gold ($Au^{79}$) and tungsten ($W^{74}$), with their dimensions listed in Table~\ref{tab:5-01}.\
These dimensions and high-resolution images of the double-slit foils are presented in 
Ref.~\cite{castle2025}.\
\begin{table}[ht]
    \centering
    \caption{The DS slit dimensions.~The thicknesses are reported to be less than~\SI{0.5}{\milli \meter}.}
    \label{tab:5-01}
    \renewcommand{\arraystretch}{1.5} % Adjust row height
    \setlength{\tabcolsep}{14pt} % Adjust column spacing
    \begin{tabular}{@{} lccr @{}} % Define column alignment
        \toprule
        \textbf{Slit no.} & \textbf{Width} ($\mu m$)& \textbf{Spacing}($\mu m$) & \textbf{ Material}  \\
        \midrule
        1& 7 $\pm$4& 25$\pm$2& W\\
        2& 8 $\pm$2 & 50$\pm$2 & Au\\
        3& 10 $\pm$2& 100$\pm$2 & Au\\
        \bottomrule
    \end{tabular}
\end{table}
The gold DSs were made using Focused Ion Beam (FIB) milling by McMaster University Photonics Research Laboratory~\cite{fib_macmaster_2021}.\
The tungsten DS was created using a laser drill purchased from Lenox Laser Inc.~\cite{lenox-laser}.\
The experimental details are described comprehensively in Ref.~\cite{yys_phd}.\

\subsubsection{X-ray Diagnostic beamline (XSR)}
The X-ray Synchrotron Diagnostic beamline at CLS is equipped with an X-ray pinhole detector set up to monitor the beam~\cite{Bergstrom2008}.\
The vertical beam size is measured as the coupling factor changes for control measures, see Fig.\ref{fig:linxsr}.
%\clearpage
\section{Results} % Data results and analysisi could be one section
\begin{figure}[tb]
    \centering
    \includegraphics[width=\linewidth]{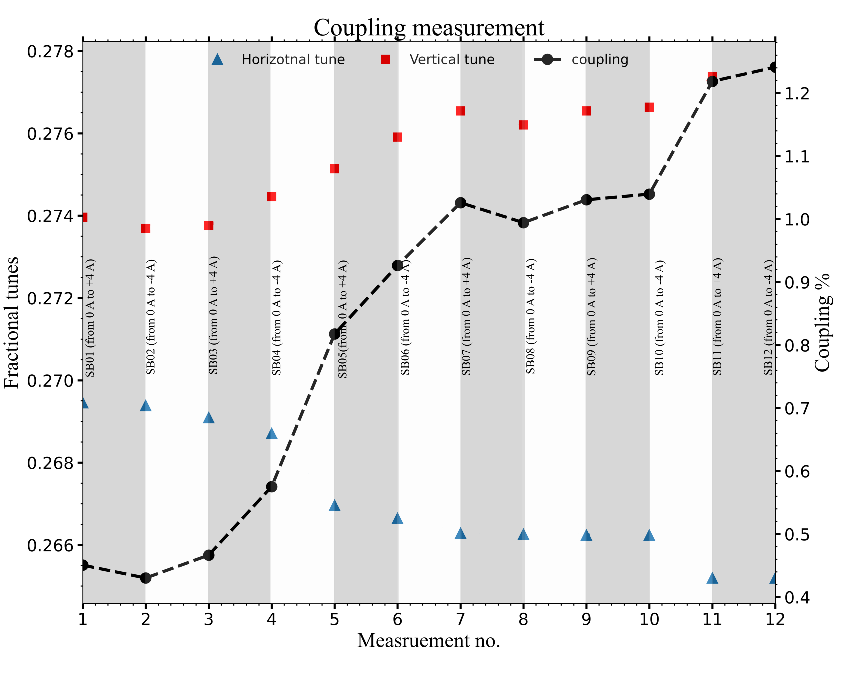}
    \caption{The closest-tune SR states at different coupling states, the region of interest for the interferometry method is the changes between step number one and two, where the coupling is reduced.}
    \label{fig:4-bbb-tbt}
\end{figure}

In this section, we present the spatial degree of coherence as a function of the electron beam coupling, an online copy is provided in Ref.~\cite{data_zenodo_2025} as well.\ 
The coupling factor $\kappa$, defined by Eq.~\ref{eq:6}, was determined using a closest-tune approach by measuring the fractional tune difference between the horizontal and the vertical planes.\
The tune-difference is a measure of the global coupling in the storage ring.\
Figure~\ref{fig:4-bbb-tbt} shows the correlation between the transverse fractional tunes and the coupling factor.\
For a selected values of the coupling, interferograms were collected on the BXDS-IVU beamline.\
The vertical beam size in XSR was collected as the control measure, and shown in Fig.~\ref{fig:linxsr}.\
The vertical fractional tune is represented by the red squares, the horizontal fractional tune by the blue triangles, and the linear coupling $\kappa$ is denoted by black line.\
The coupling is adjusted by modifying the skew quadrupoles and observing the closest approach between the vertical and horizontal fractional tune values.\
The tunes were measured by driving the beam with a ramped frequency and identifying the resonance peaks in the response signals.\
This demonstrates our ability to control the global coupling in the storage ring.\

\section{Analysis}
\begin{figure}[htbp]
    \centering
    \includegraphics[width=\linewidth]{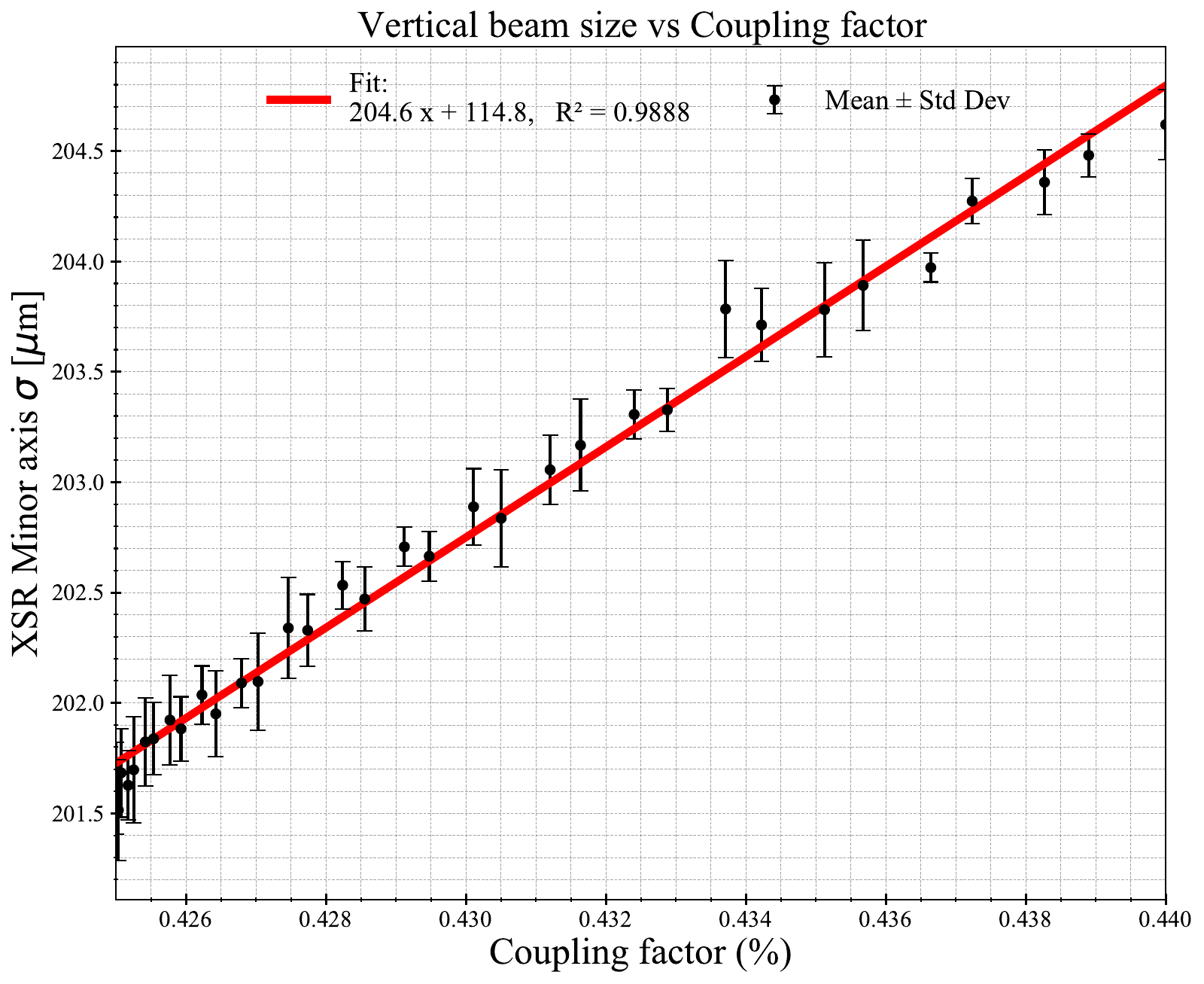}
    \caption{Vertical source size as a function of coupling factor.}
    \label{fig:linxsr}
\end{figure}

\begin{figure}[htbp]
    \centering
    \includegraphics[width=\linewidth]{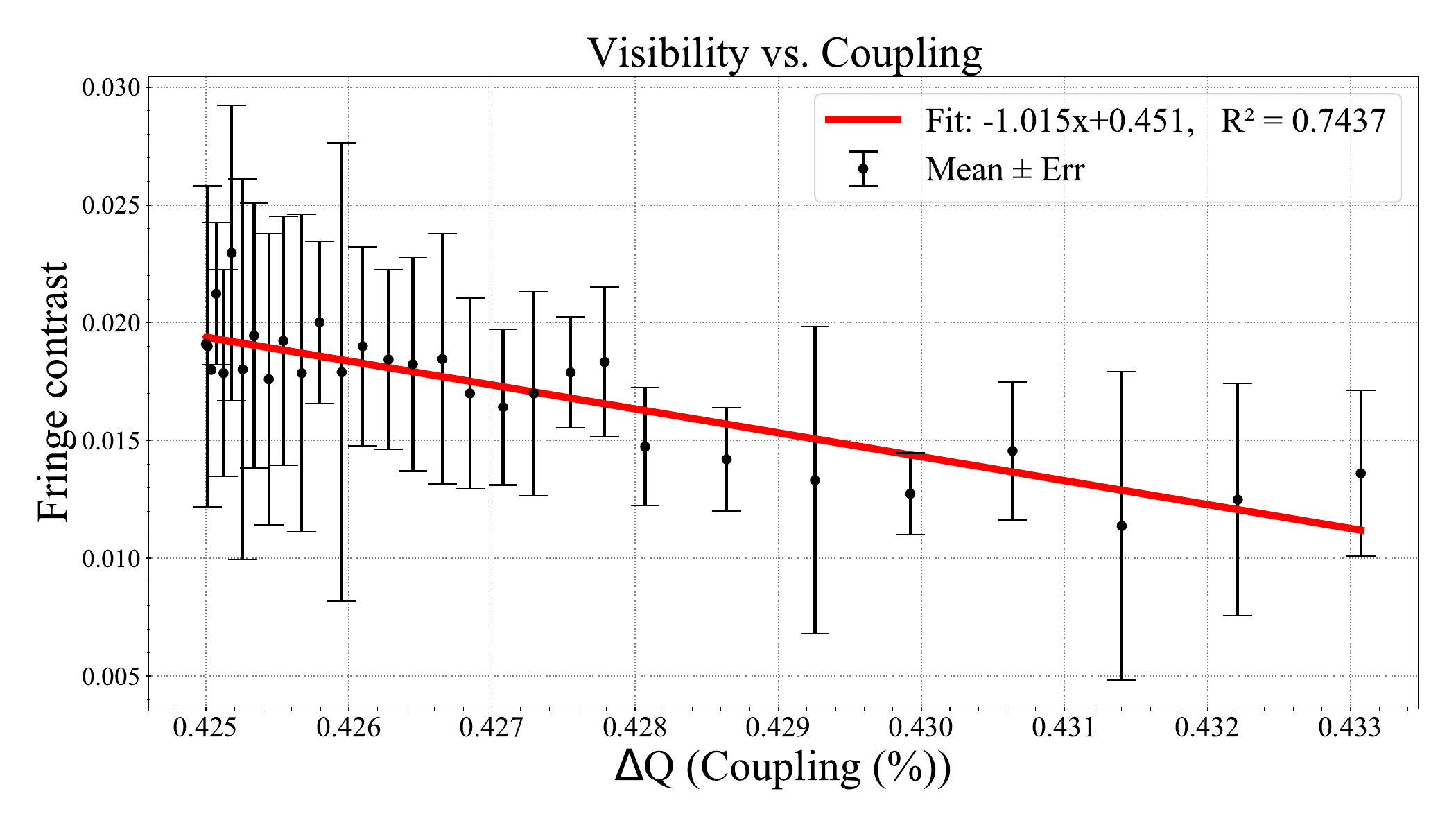}
    \caption{Visibility as a function of linear betatron coupling $\kappa$ with statistical error.}
    \label{fig:correctr_viz}
\end{figure}

The XSR diagnostic beamline shows consistency with the LOCO predictions as the vertical beam size increases linearly with changing the coupling factor defined by the closest tune approach, see Fig.~\ref{fig:linxsr}.
\FloatBarrier
The plot in \hfigref{fig:correctr_viz} shows the mean value for the visibility data as a function of coupling factor.\
The coupling varies with changes in the skew quadrupole currents, leading to a decrease and subsequent increase in the vertical source size.\
The maximum and minimum visibilities are~\SI{0.0225}{} and~\SI{0.012}{}, respectively.\
\begin{table}[htbp]
    \centering
    \caption{A list of measured visibility with the uncertainty propagation, the analyzed degree of coherence at photon energy \SI{6.99}{\kilo eV}, and the van~Cittert-Zernike theorem predictions of beam height.~The DS is positioned at~\SI{50}{\meter} away from the source point.}    \renewcommand{\arraystretch}{1.4} % Adjust row height
    \setlength{\tabcolsep}{1pt} % Adjust column spacing
    \begin{tabular}{@{} lcc @{}} % Define column alignment
        \toprule
       % \rowcolor{gray!20}
       \small \textbf{Visibility}&\small \textbf{Degree of coherence } & \small \textbf{Vertical beam size}  \\
        \small (no units) &\small  (\SI{}{\micro \meter}) & \small (\SI{}{\micro \meter})   \\
        \midrule
        0.0125~$\pm$~0.0006 & 10.75~$\pm$~0.09 & 80.1~$\pm$~0.2 \\
        \hline
        0.0147~$\pm$~0.0006 & 10.96~$\pm$~0.09 & 78.6~$\pm$~0.3  \\       \hline
        0.0179~$\pm$~0.0007 &  11.22~$\pm$~0.10  & 76.8~$\pm$~0.3 \\
        \hline
        0.0230~$\pm$~0.0008 & 11.59~$\pm$~0.10  &  74.3~$\pm$~0.4 \\     
        \bottomrule
    \end{tabular}
    \label{tab:6-viz-size3}
\end{table}

The data were then fitted with a weighted linear regression to analyze the visibility trend as a function of the coupling factor.
The results indicate that visibility increases as the linear betatron coupling decreases, as shown in \hfigref{fig:correctr_viz}; this trend aligns with the theoretical predictions for the first order spatial degree of coherence.\
The first order spatial degree of coherence measured by synchrotron radiation interferometry has an inverse relation with the coupling factor measured by the closest-tune approach.\
A selection of data points is provided in Table~\ref{tab:6-viz-size3} for reference.
\FloatBarrier
%\clearpage
\section{Conclusion}
This study demonstrates the application of X-ray double-slit interferometry to serve as a diagnostic technique for measuring the spatial degree of coherence of synchrotron radiation.\
This was demonstrated on the BXDS-IVU beamline at CLS with X-ray beam energy of~\SI{7}{keV}.\
By systematically changing the coupling factor in the storage ring and observing its impact on the degree of spatial coherence, we were able to assess the relation between the visibility and coupling.\
The use of skew quadrupoles in the DBA cell, along with precise interferometry measurements, provided valuable insights into optimizing the transverse degree of coherence from visibility of \SI{0.0125}{} to \SI{0.0230}{} for the slit spacing of~\SI{50}{\micro \meter}.\

The experimental results contribute to our understanding of beam dynamics in synchrotron light sources and have potential applications in beamline design and operational optimization.\
This work lays the foundation for further studies to control of the degree coherence and its impact on synchrotron radiation experiments.\

\section{Acknowledgment}
This research was supported by the Natural Sciences and Engineering Research Council of Canada (NSERC). The beamtimes were a courtesy of the Canadian Light Source through the University of Saskatchewan.~The authors acknowledge the resources provided by the The Biomedical Imaging and Therapy Facility (BMIT) staff, the control room operator group and the AOD team; Dr.~Ward~Wurtz and Dr.~Melissa~Ratzlaff.

\printcredits

\bibliographystyle{unsrt}
\bibliography{YYS-ref-initials}

@phdthesis{yys_phd,
  author = {Y. Y. Sigari},
  title     = {Coherence Control and Measurement at the CLS Using X-ray Interferometry},
  school    = {University of Saskatchewan},
  year      = {2025},
  address   = {Saskatoon, Canada}
}

@article{Takayama:el3109,
  author = {Y. Takayama and R. Z. Tai and T. Hatano and T. Miyahara and W. Okamoto and Y. Kagoshima},
  title     = {Measurement of the coherence of synchrotron radiation},
  journal   = {Journal of Synchrotron Radiation},
  year      = {1998},
  volume    = {5},
  number    = {3},
  pages     = {456--458},
  month     = may,
  doi       = {10.1107/S0909049597013010},
  url       = {https://doi.org/10.1107/S0909049597013010},
  publisher = {International Union of Crystallography}
}

@article{Chushkin:ro5044,
author = "Chushkin, Y. and Zontone, F.",
title = "{Prospects for coherent X-ray diffraction imaging at fourth-generation synchrotron sources}",
journal = "IUCrJ",
year = "2025",
volume = "12",
number = "3",
pages = "280--287",
month = "May",
doi = {10.1107/S2052252525001526},
url = {https://doi.org/10.1107/S2052252525001526},
}

@article{YousefiSigari:2021hwg,
    author = "Sigari, Y. and Bertwistle, D. and Boland, M. J.",
    title ={{Vertical Phase Space Measurement Progress at {C}anadian {L}ight {S}ource}},
    doi = "10.18429/JACoW-IPAC2021-MOPAB310",
    journal = "JACoW",
    volume = "IPAC2021",
    pages = "MOPAB310",
    year = "2021"
}

@article{castle2025, 
    author = "Castle, R. and Appathurai, N. and  Simonson, N. and Sigari, Y. and Boland, M.~J. and He, F. and Karunakaran, C. and Wang, J. and Moreno, B.~D. and Kuppili, V.~S.~C. ",
    title = {{Investigating the limits of hard X-ray coherence length measurement employing Young’s double slit experiment}},
    journal = "Scientific Reports" ,
    volume = "15",
    doi = "https://doi.org/10.1038/s41598-025-03295-y",
    year = "2025"
}

@article{canlight_CutlerJeffrey2017Tbli,
issn = {2412-382X},
journal = {Quantum beam science},
pages = {4},
volume = {1},
publisher = {MDPI AG},
number = {1},
year = {2017},
title = {The brightest light in {C}anada: The {C}anadian {L}ight {S}ource},
copyright = {Copyright 2021 Elsevier B.V., All rights reserved.},
language = {eng},
author = {J. Cutler and D. Chapman and L. Dallin and R. Lamb},
}

@article{4thgen_chal_2014,
  author = {M. Borland and G. Decker and L. Emery and V. Sajaev and Y. Sun and A. Xiao},
  title     = {Lattice design challenges for fourth-generation storage-ring light sources},
  journal   = {Journal of Synchrotron Radiation},
  year      = {2014},
  volume    = {21},
  number    = {5},
  pages     = {912--936},
  doi       = {10.1107/S1600577514015203},
  url       = {https://doi.org/10.1107/S1600577514015203},
  publisher = {International Union of Crystallography}
}

@article{cdi_gale_infotracmisc_A769864464,
  author = {C. L. Walsh and P. Tafforeau and W. L. Wagner and D. J. Jafree and A. Bellier and C. Werlein and M. P. Kühnel and E. Boller and S. Walker-Samuel and J. L. Robertus and D. A. Long and J. Jacob and S. Marussi and E. Brown and N. Holroyd and D. D. Jonigk and M. Ackermann and P. D. Lee},
  title   = {Imaging intact human organs with local resolution of cellular structures using hierarchical phase-contrast tomography},
  journal = {Nature Methods},
  volume  = {18},
  number  = {12},
  pages   = {1532--1541},
  year    = {2021},
  doi     = {10.1038/s41592-021-01317-x},
  url     = {https://doi.org/10.1038/s41592-021-01317-x},
  language= {English},
  publisher= {Springer Nature}
}

@article{Zernike1938,
   author = {F. Zernike},
   doi = {https://doi.org/10.1016/S0031-8914(38)80203-2},
   issn = {0031-8914},
   issue = {8},
   journal = {Physica},
   pages = {785-795},
   title = {The concept of degree of coherence and its application to optical problems},
   volume = {5},
   url = {https://www.sciencedirect.com/science/article/pii/S0031891438802032},
   year = {1938}
}

@misc{2021MATLABR2021a,
    title = {{MATLAB version 9.10.0.1613233 (R2021a)}},
    year = {2021},
    address = {Natick, Massachusetts}
}

@article{vanCittert1934,
   author = {P. H. v. Cittert},
   doi = {https://doi.org/10.1016/S0031-8914(34)90026-4},
   issn = {0031-8914},
   issue = {1},
   journal = {Physica},
   pages = {201-210},
   title = {Die Wahrscheinliche Schwingungsverteilung in Einer von Einer Lichtquelle Direkt Oder Mittels Einer Linse Beleuchteten Ebene},
   volume = {1},
   url = {https://www.sciencedirect.com/science/article/pii/S0031891434900264},
   year = {1934}
}

@article{Shin2021,
  author = {S. Shin},
  title     = {New era of synchrotron radiation: fourth-generation storage ring},
  journal   = {AAPPS Bulletin},
  year      = {2021},
  volume    = {31},
  number    = {1},
  pages     = {21},
  doi       = {10.1007/s43673-021-00021-4},
  url       = {https://doi.org/10.1007/s43673-021-00021-4},
  issn      = {2522-4120},
  publisher = {Springer Nature on behalf of the Association of Asia Pacific Physical Societies}
}

@article{Bergstrom2008,
author = {Bergstrom, J.~C. and Vogt, J.~M.},
doi = {10.1016/j.nima.2008.01.080},
journal = {Nuclear Instruments and Methods in Physics Research Section A: Accelerators, Spectrometers, Detectors and Associated Equipment},
month = {Jan},
pages = {441--457},
title = {{The X-ray diagnostic beamline at the Canadian Light Source}},
volume = {587},
year = {2008}
}

@article{bemaenvelope_ohmi_1994,
  title = {From the beam-envelope matrix to synchrotron-radiation integrals},
  author = {Ohmi, K. and Hirata, K. and Oide, K.},
  journal = {Physical Review E},
  volume = {49},
  issue = {1},
  pages = {751--765},
  numpages = {0},
  year = {1994},
  month = {Jan},
  publisher = {American Physical Society},
  doi = {10.1103/PhysRevE.49.751},
  note = {https://link.aps.org/doi/10.1103/PhysRevE.49.751}
}

@misc{fib_macmaster_2021,
  title  = {{Photonics Research Laboratories Home (Focused Ion Beam)}},
  author = {{McMaster University}},
  year   = {2021},
  note   = {https://physics.mcmaster.ca/optics/FIB.html},
  url    = {https://physics.mcmaster.ca/optics/FIB.html}
}

@misc{lenox-laser,
  title  = {{Lenox Laser Inc.}},
  year   = {2021},
  note   = {https://lenoxlaser.com/ , Accessed: 10 July 2021},
  url    = {https://lenoxlaser.com/}
}

@techreport{detector_man_2018,
    author = "MAN-11421-1811-0347-A",
    title = "Monochromatic beam microscope- single objective user manual",
    institution = "Canadian Light Source",
    year = "2018/11/05",
}

@article{couple_WURTZ_2018,
title = {{Coupling control and optimization at the Canadian Light Source}},
journal = {Nuclear Instruments and Methods in Physics Research Section A: Accelerators, Spectrometers, Detectors and Associated Equipment},
volume = {892},
pages = {1-9},
year = {2018},
issn = {0168-9002},
doi = {https://doi.org/10.1016/j.nima.2018.02.082},
url = {https://www.sciencedirect.com/science/article/pii/S0168900218302626},
author = {W. A. Wurtz}
}

@inproceedings{canmag_dallin_2003,
    author = {L. Dallin and D. Lowe and J. Swirsky},
    title = {{Canadian Light Source Magnets}},
    booktitle = {Proc. PAC'03},
    pages = {2195--2197},
    paper = {WPAB068},
    venue = {Portland, OR, USA, May 2003},
    publisher = {JACoW Publishing, Geneva, Switzerland},
    url = {https://jacow.org/p03/papers/WPAB068.pdf},
    year = {2003},
    language = {english}
}

@INPROCEEDINGS{cls_design_dallin_2003,
 author = {Dallin, L. and Blomqvist, I. and Jong, M.~D. and Lowe, D. and Silzer, M.},
  title     = {{The Canadian Light Source}},
  booktitle = {Proceedings of the 2003 Particle Accelerator Conference (PAC 2003)},
  year      = {2003},
  volume    = {1},
  pages     = {220--223},
  doi       = {10.1109/PAC.2003.1288884},
  address   = {Portland, Oregon, USA},
  publisher = {IEEE}
}

@inproceedings{mitsuhashi_lhc_vdbl,
  author = {G. Trad and E. Bravin and A. Goldblat and S. Mazzon and F. Roncarolo and T. Mitsuhashi},
  title     = {{Performance of the Upgraded Synchrotron Radiation Diagnostics at the LHC}},
  booktitle = {Proceedings of the International Particle Accelerator Conference (IPAC 2016)},
  year      = {2016},
  address   = {Busan, Korea},
  pages     = {215--218},
  doi       = {10.18429/JACoW-IPAC2016-THPOR035},
  url       = {https://doi.org/10.18429/JACoW-IPAC2016-THPOR035}
}

@article{Mitsuhashi2016,
   author = {Mitsuhashi, T. and Oide, K. and Zimmermann, F.},
   doi = {doi:10.18429/JACoW-IPAC2016-MOPMB022},
   journal = {Proc. of International Particle Accelerator Conference (IPAC16), Busan, Korea},
   month = {8},
   pages = {133-136},
   title = {{Conceptual Design for SR Monitor in the FCC Beam Emittance Size Dagnostic}},
   url = {http://jacow.org/ipac2016/papers/mopmb022.pdf},
   year = {2016},
}

@article{bxds_diaz_2014,
    author = {B. Diaz and e. al.},
    title = "{Undulator beamline of the Brockhouse sector at the Canadian Light Source}",
    journal = {Review of Scientific Instruments},
    volume = {85},
    number = {8},
    pages = {085104},
    year = {2014},
    month = {08},
    issn = {0034-6748},
    doi = {10.1063/1.4890815},
    url = {https://doi.org/10.1063/1.4890815},
    eprint = {https://pubs.aip.org/aip/rsi/article-pdf/doi/10.1063/1.4890815/9979902/085104\_1\_online.pdf},
}

@article{Chasman_1975_PreliminaryDO,
  title={Preliminary Design of a Dedicated Synchrotron Radiation Facility},
  author = {R. W. Chasman and G. K. Green and E. M. Rowe},
  journal={IEEE Transactions on Nuclear Science},
  year={1975},
  volume={22},
  pages={1765-1767},
  url={https://api.semanticscholar.org/CorpusID:21358102}
}

@techreport{acctool_terebilo_2001,
    author = {A. Terebilo},
    title = {Accelerator Toolbox for {MATLAB}},
    number      = {SLAC-PUB-8732},
    institution = {SLAC},
    year = {2001},
    doi = {10.2172/784910}
}

@book{minty2003,
  author = {M. G. Minty and F. Zimmermann},
  title     = {Measurement and Control of Charged Particle Beams},
  publisher = {Springer Nature},
  year      = {2003},
  doi       = {10.1007/978-3-662-08581-3_13},
  pages ={56-62},
  address   = {Berlin, Germany}
}

@article{loco_safranek2009,
  author = {J. Safranek},
  title     = {{Linear Optics from Closed Orbits ({L}{O}{C}{O}): An Introduction}},
  journal   = {ICFA Beam Dynamics Newsletter},
  volume    = {44},
  pages     = {43--49},
  year      = {2009},
  month     = {June},
  url       = {https://www.osti.gov/biblio/957449},
}

@article{young-1804-interference,
  author = {T. Young},
  title     = {{I. The Bakerian Lecture. Experiments and calculations relative to physical optics}},
  journal   = {Philosophical Transactions of the Royal Society of London},
  volume    = {94},
  pages     = {1--16},
  year      = {1804},
  doi       = {10.1098/rstl.1804.0001},
  url       = {https://royalsocietypublishing.org/doi/abs/10.1098/rstl.1804.0001},
  eprint    = {https://royalsocietypublishing.org/doi/pdf/10.1098/rstl.1804.0001}
}

@book{Born_Wolf,
  author = {M. Born and E. Wolf},
  title     = {Principles of Optics},
  edition   = {7th},
  publisher = {Cambridge University Press},
  year      = {1999},
  address   = {Cambridge, UK},
  isbn      = {978-0-521-64222-4}
}

@misc{data_zenodo_2025,
  author       = {Sigari, Y.~Y. and Boland, M.~J.},
  title        = {{X-ray interferometer and pinhole data from the CLS BXDS beamline during storage ring coupling scans}},
  publisher    = {Zenodo},
  year         = {2025},
  note          = { 10.5281/zenodo.16416212},
  url          = {https://doi.org/10.5281/zenodo.16416212}
}

\end{document}